\documentclass[aps,pra,twocolumn,floatfix,10pt,showpacs,superscriptaddress,longbibliography]{revtex4-1}

\usepackage{graphicx}
\usepackage{verbatim}
\usepackage{comment}
\usepackage{amssymb}
\usepackage{enumerate}
\usepackage{psfrag}
\usepackage{pstricks}
\usepackage{color}
\usepackage{hyperref}
\usepackage{epstopdf}
\usepackage{MnSymbol}
\usepackage{float}
\definecolor{navyblue}{RGB}{46,48,146}

\hypersetup{
  colorlinks  =true, 
  urlcolor    =navyblue, 
  linkcolor   =navyblue, 
  citecolor   =navyblue 
}

\emergencystretch=\maxdimen
\begin{document}

\title{Machine Learning Phases of Strongly Correlated Fermions}

\author{Kelvin Ch'ng}
\affiliation{Department of Physics and Astronomy, San Jos\'{e} State University, San Jos\'{e}, CA 95192, USA}
\author{Juan Carrasquilla}
\affiliation{Perimeter Institute for Theoretical Physics, Waterloo, Ontario N2L 2Y5, Canada}
\author{Roger G. Melko}
\affiliation{Perimeter Institute for Theoretical Physics, Waterloo, Ontario N2L 2Y5, Canada}
\affiliation{Department of Physics and Astronomy, University of Waterloo, Ontario N2L 3G1, Canada}
\author{Ehsan Khatami}
\affiliation{Department of Physics and Astronomy, San Jos\'{e} State University, San Jos\'{e}, CA 95192, USA}

\begin{abstract}
Machine learning offers an unprecedented perspective for the problem of classifying phases in condensed matter physics. We employ neural-network machine learning techniques to distinguish finite-temperature phases of the strongly correlated fermions on cubic lattices. We show that a three dimensional convolutional network trained on auxiliary field configurations produced by quantum Monte Carlo simulations of the Hubbard model can correctly predict the magnetic phase diagram of the model at the average density of one (half filling). We then use the network, trained at half filling, to explore the trend in the transition temperature as the system is doped away from half filling. This transfer learning approach predicts that the instability to the magnetic phase extends to at least 5\% doping in this region. Our results pave the way for other machine learning applications in correlated quantum many-body systems.

\end{abstract}

\maketitle

\section{Introduction}

\begin{figure}[b]
\centerline {\includegraphics*[width=3.3in]{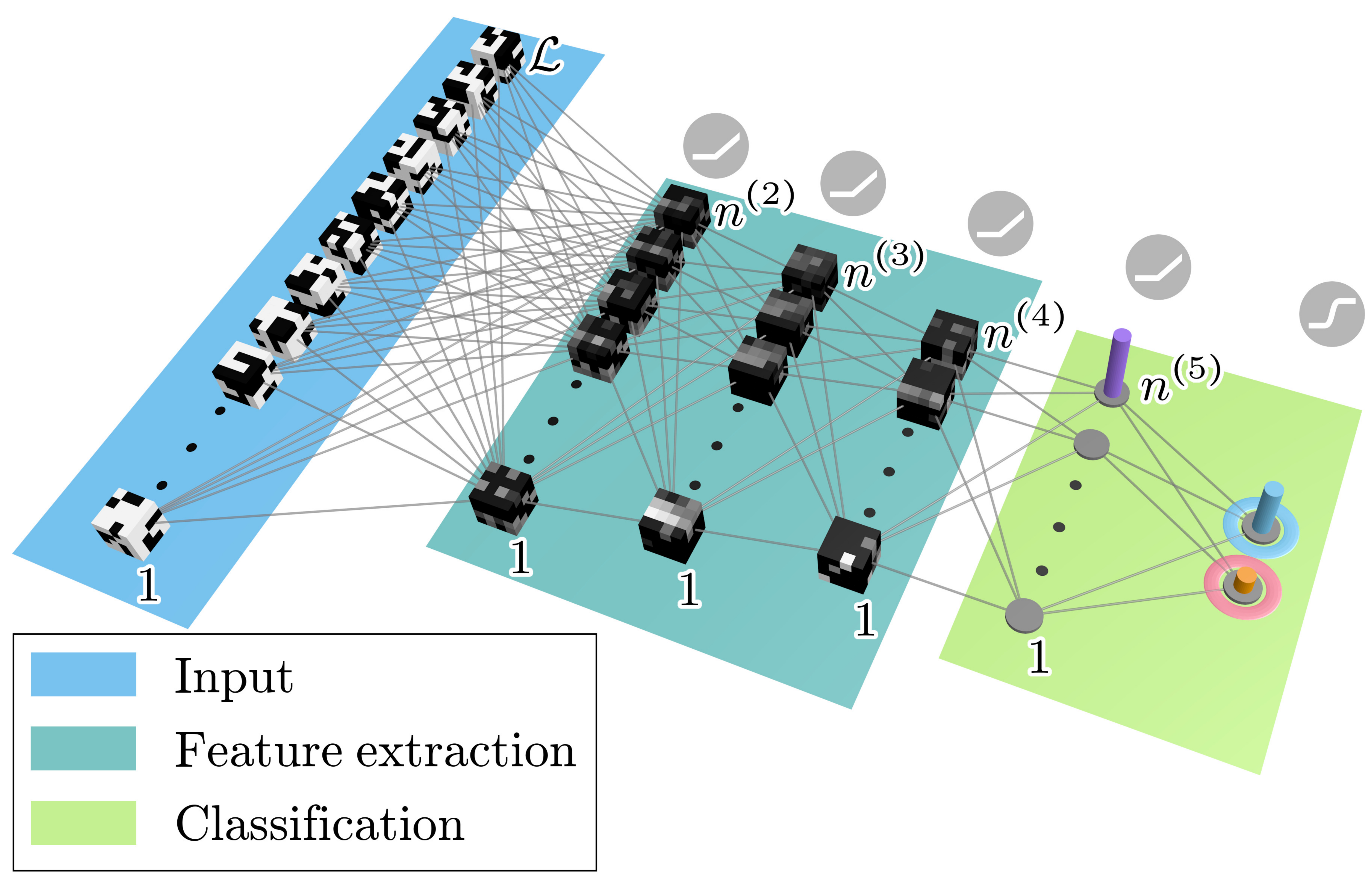}}
\caption{
Architecture of the 3D convolutional neural network used to obtain $T_{N}$ for the 3D Hubbard model. For input, we use the auxiliary field configurations in a four-dimensional grid, three spatial dimensions of size 4 (total of $N=4^3$ sites), and one imaginary time dimension of size $\mathcal{L}=200$. Numbers of volumetric feature maps in the hidden-feature extraction layers are $n^{(2)}=32$, $n^{(3)}=16$, and $n^{(4)}=8$. Here, $n^{(5)}=8$ the fully connected layer. During training, dropout regularization with a rate of 0.5 was used to mitigate overtraining. To classify the input system as ordered or unordered, each of the eight fully connected neurons is connected to each of the two readout neurons using the softmax function as a neural activation function. The output neuron with the highest probability represents the activated neuron.
\label{fig:network}}
\end{figure}

The various modern architectures of neural networks consisting of multiple layers and neuron types (see Fig.~\ref{fig:network} for an example) can be {\em trained} to classify, with a high degree of accuracy, intricate sets of labeled data [1]. The data, e.g., a series of handwritten digits, are fed to the network input layer, and the outcome, read at the output layer, are neuron activations corresponding to the different digits. Common to most algorithms involving neural networks is the training procedure, which is an optimization problem where the free parameters associated with connections between neurons in adjacent layers and their biases (additive constants) are slowly adjusted until a high classification accuracy is attained. Embodied in the study of quantum and classical statistical mechanics are the many-body states, which can be understood as immense data sets associated with the equilibrium state of the system, and over which machine learning techniques can be naturally applied. Early applications of machine learning ideas in condensed matter physics focused on their connection to renormalization group methods~\cite{p_mehta_14}, obtaining the Green's function of the Anderson impurity model model~\cite{l_arsenault_14}, categorizing real materials~\cite{a_kusne_14,s_kalinin_15,l_ghiringhelli_15,s_schoenholz_16}, or learning ground states and thermodynamics of many-body systems~\cite{Carleo2016,Torlai2016}. Tensor-network representations of quantum states have also been proposed recently as a powerful tool for supervised learning~\cite{Stoudenmire2016}.    

Recently, neural-network machine learning algorithms have been successfully adopted to distinguish phases of matter in classical Ising-type models, effectively locating critical temperatures at which transitions between phases take place~\cite{j_carrasquilla_16,Wang2016}. Two of the authors found that by using a simple network consisting of only one hidden layer, one can predict the transition temperature with up to 99\% accuracy for two-dimensional (2D) Ising models, solely based on spin configurations generated by Monte Carlo simulations and in the absence of any information about the underlying lattice or the order parameter [11].

The extension of the technique to quantum mechanical systems is less straightforward, as the quantum Monte Carlo simulations of interacting particles involve an additional dimension associated with imaginary time in the path integral formalism at finite temperatures, or the projection parameter for ground-state calculations; quantum fluctuations can distort the easily recognizable picture of spin configurations in the ordered phase of the classical system and therefore significantly affect the training process of the neural network. Here, using quantum Monte Carlo simulations of the Hubbard model of strongly correlated fermions on cubic lattices and convolutional neural networks (CNNs), we show that one can successfully classify finite-temperature phases of quantum systems and estimate transition temperatures with a reasonable degree of accuracy on relatively small lattice sizes.

\section{Model}

The Fermi-Hubbard Hamiltonian~\cite{n_mott_49,j_hubbard_63} in the particle-hole
invariant form is expressed as

\begin{equation}
H=-t\sum_{\left<ij\right> \sigma}c^{\dagger}_{i\sigma}
c^{\phantom{\dagger}}_{j\sigma} + U\sum_i \left( n_{i\uparrow}-\frac{1}{2} \right) \left(n_{i\downarrow}-\frac{1}{2} \right)
-\mu\sum_{i\sigma} n_{i\sigma},
\label{eq:H}
\end{equation}

\noindent
where $c^{\phantom{\dagger}}_{i\sigma}$ ($c^{\dagger}_{i\sigma}$) annihilates (creates) a fermion with spin $\sigma$ on site $i$,
$n_{i\sigma}=c^{\dagger}_{i\sigma} c^{\phantom{\dagger}}_{i\sigma}$ is the number operator, $U$ is the onsite Coulomb interaction,  $\left< \cdot \cdot \right>$ denotes nearest neighbors, $t$ is the corresponding hopping integral, and $\mu$ is the chemical potential. Here, $\mu=0$ corresponds to the half-filled model (average density of one fermion per site, $n=1$). We set $t=1$ as the unit of energy and consider the model on three-dimensional (3D) cubic lattices.

The 3D model at half filling realizes a finite-temperature transition to the antiferromagnetic N\'{e}el phase for any $U>0$, analogous to the magnetic ordering in the 2D classical Ising model. The transition temperature $T_N$, which is relatively well known from the analysis of the staggered spin structure factor, or the staggered susceptibility~\cite{r_scalettar_89,r_staudt_00,p_kent_05,t_paiva_11,e_kozik_13,d_hirschmeier_15,e_khatami_16}, is a nonmonotonic function of the interaction strength; it increases rapidly with increasing $U$ in the weak-coupling regime ($U \lesssim 8$), a result that can be captured using the random phase approximation~\cite{r_scalettar_89}, and decreases at large $U$. In the strong-coupling regime ($U \gtrsim 12$), the half-filled model can be effectively described by the antiferromagnetic (AFM) Heisenberg model, whose exchange constant, and hence, N\'{e}el temperature, is proportional to $1/U$~\cite{a_sandvik_98}.

\section{Method}

Our goal here is to train a CNN to identify finite-temperature phase boundaries of the Hubbard model. We utilize the determinantal quantum Monte Carlo (DQMC)~\cite{r_blankenbecler_81}, which reduces the numerical evaluation of the observables of the Fermi-Hubbard model to a stochastic averaging over a set of discrete auxiliary fields extending in space and along an {\em imaginary time} dimension. The spin correlations of the model can be written directly in terms of the correlations in our particularly chosen auxiliary field (see Appendix~$\hyperref[app:Appendix_A]{\text{A}}$), rendering it an obvious choice to be used in the identification of magnetic phases through machine learning, although a previous attempt by including two of the authors has not been successful~\cite{p_broecker_16}. The training is performed using the field configurations generated during DQMC simulations in a range of temperatures around one or two critical points. The objective is to use the trained network to map out the entire phase boundary associated with the same critical phenomenon by varying the parameters driving the transition and generating {\em test} data sets of the field configurations. In this work, we focus on the magnetic properties of the Hubbard model.

We use a 3D CNN, originally developed for human action recognition in videos~\cite{s_ji_2013}, implemented in Tensorflow~\cite{tensorflow}. Convolutions are designed to return information about spatial dimension and locality to the simpler idea of a fully connected feed-forward neural network. In our case, the three spatial dimensions of the cubic lattice are treated with the convolution,while slices in the fourth imaginary time axis are used as different filter channels~\cite{Nielsen2016}. The network architecture for $N=4^3$ is shown in Fig.~\ref{fig:network}. We use three or four hidden layers, depending on the spatial size of the system, for feature extraction, followed by a fully connected layer before the output layer. The optimal number of neurons in each layer (resulting in the largest accuracy) for $N=8^3$ is found using a Monte Carlo optimization procedure (see Appendix~$\hyperref[app:Appendix_B]{\text{B}}$).

\section{results}

To benchmark our results and validate our approach, we start with the 3D Hubbard model at half filling and explore the accuracy with which we can predict the N\'{e}el phase boundary in the temperature-interaction space.We train the network to distinguish (by activating the corresponding output neuron) spin configurations belonging to the ordered phase ($T<T_N$) from those of the unordered high-temperature phase (see Appendix~$\hyperref[app:Appendix_C]{\text{C}}$). The approximately 80 000 labeled configurations at various temperatures around $T_{N}$ are generated through DQMC simulations for two interaction strengths, $U=5$ and 16, one in the weak-coupling and one in the strong-coupling regime, and shuffled before they are used in the training. The trained network is then used to classify other configurations as the temperature is varied across the estimated critical values for other values of $U$ between 4 and 16.

\begin{figure}[t]
\centerline {\includegraphics*[width=3.3in]{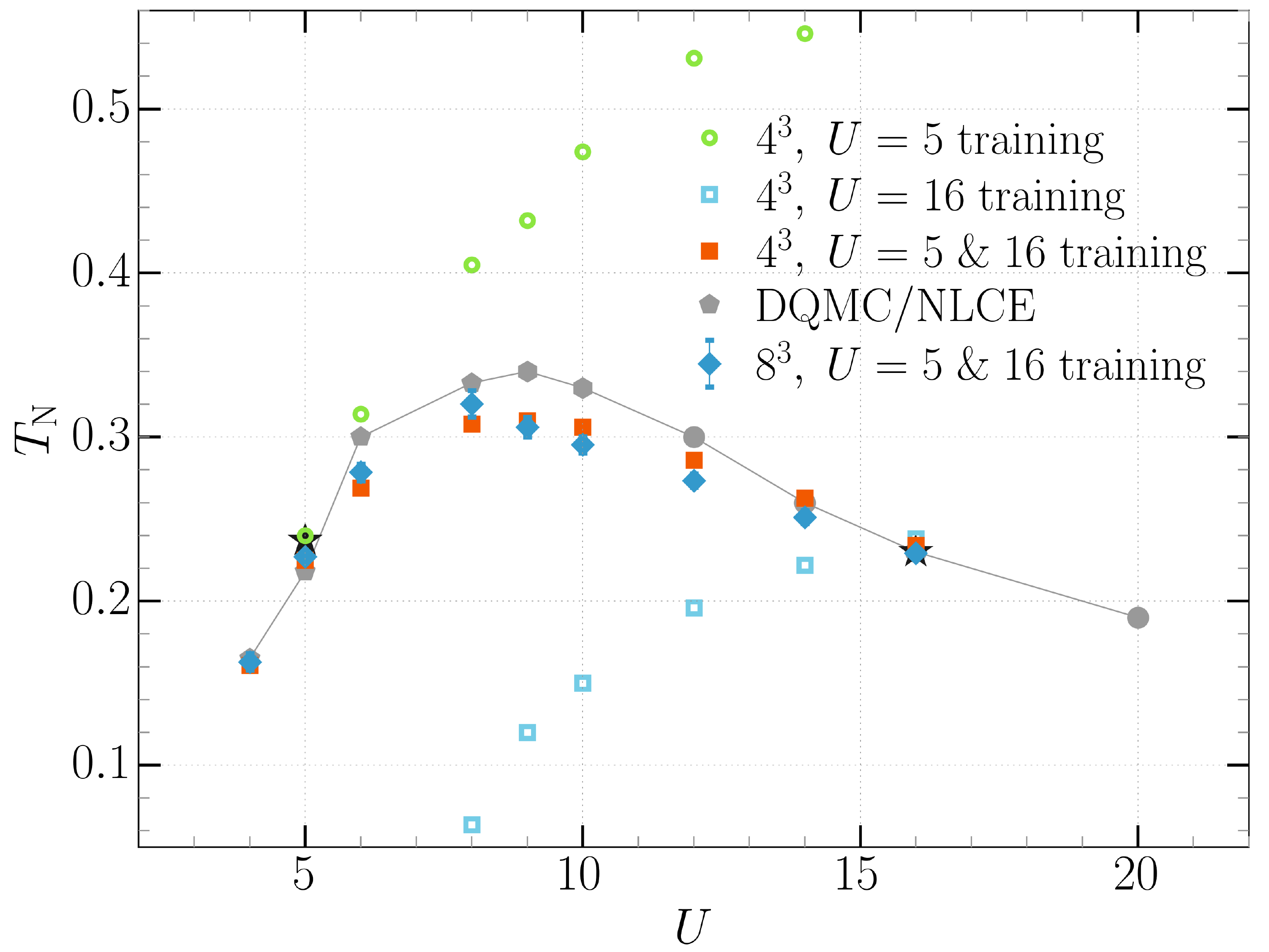}}
\caption{Prediction of the N\'{e}el transition temperature by the neural network. Using the auxiliary spin configurations, the network is trained separately at $U=5$ and $U=16$ for $N=4^3$, and simultaneously at $U=5$ and 16 for $N=4^3$ and $N=8^3$. The error bars are the standard error of the mean of six different classifications using CNNs that were trained starting from different random weights and biases. The critical temperatures used for the training of the network with $N=4^3$ are shown as stars (see text). Grey filled symbols are the estimates for $T_{N}$ in the thermodynamic limit from DQMC and NLCE simulations. Grey pentagons, hexagons, and circles for weak-, intermediate-, and strong-coupling regimes are taken from Refs.~\cite{e_kozik_13, t_paiva_11, e_khatami_16}, respectively. The solid line is a guide to the eye.
\vspace{-0.1in}
\label{fig:TN}}
\end{figure}

The results for two system sizes $N=4^3$ and $N=8^3$ are shown in Fig.~\ref{fig:TN} as full squares and diamonds. Figure~\ref{fig:TN} also summarizes $T_{N}$ in the thermodynamic limit from recent unbiased studies~\cite{t_paiva_11,e_kozik_13,e_khatami_16}. One can see that, despite notable disagreements at intermediate values of $U$, remarkably, the network can predict the nontrivial shape of the phase boundary with a reasonable degree of accuracy even with a system size of $N=4^3$. The results for $N=8^3$ show the same trend. However, we find that in the strong-coupling regime, the latter are smaller than $T_N$ obtained with $N=4^3$. This counterintuitive behavior is likely rooted in the lack of precise knowledge of the critical temperature at $U=16$, and the sensitivity of CNNs' predictions to our choice of $T_{N}$ during training as explained below. For each system size, we train six different CNNs with the same architecture but with different initial random weights and biases (see Appendix~$\hyperref[app:Appendix_C]{\text{C}}$ for training details), and we show the average $T_{N}$ over different CNNs to make sure results are not biased towards any particular training. The error bars are slightly larger for $N=8^3$ as the neural network contains a significantly larger number of parameters in this case.

We note that $T_N$ is not well defined for finite clusters, which can lead to uncertainties in the labeling of the configurations in the training process. However, the {\em exact} value of $T_N$ is, in principle, not required for the training. One can omit configurations from the training data for an arbitrary temperature range in the proximity of the estimated critical value, e.g., the temperature at which the correlations reach the linear size of the cluster, at the expense of losing some predicting accuracy by the network after the training. Here, we take advantage of the fact that finite-size errors are small in the strong-coupling regime and use $T_N=0.23$ for $U=16$, obtained for the thermodynamic limit~\cite{e_khatami_16}, for both cluster sizes. For $U=5$, however, we use the approximate critical temperatures $T_N=0.24$ and $0.25$  for $N=4^3$ and $8^3$, respectively, from the analysis of the Binder ratio for the order parameter~\cite{e_kozik_13}. 

\begin{figure}[t]
\centerline {\includegraphics*[width=3.3in]{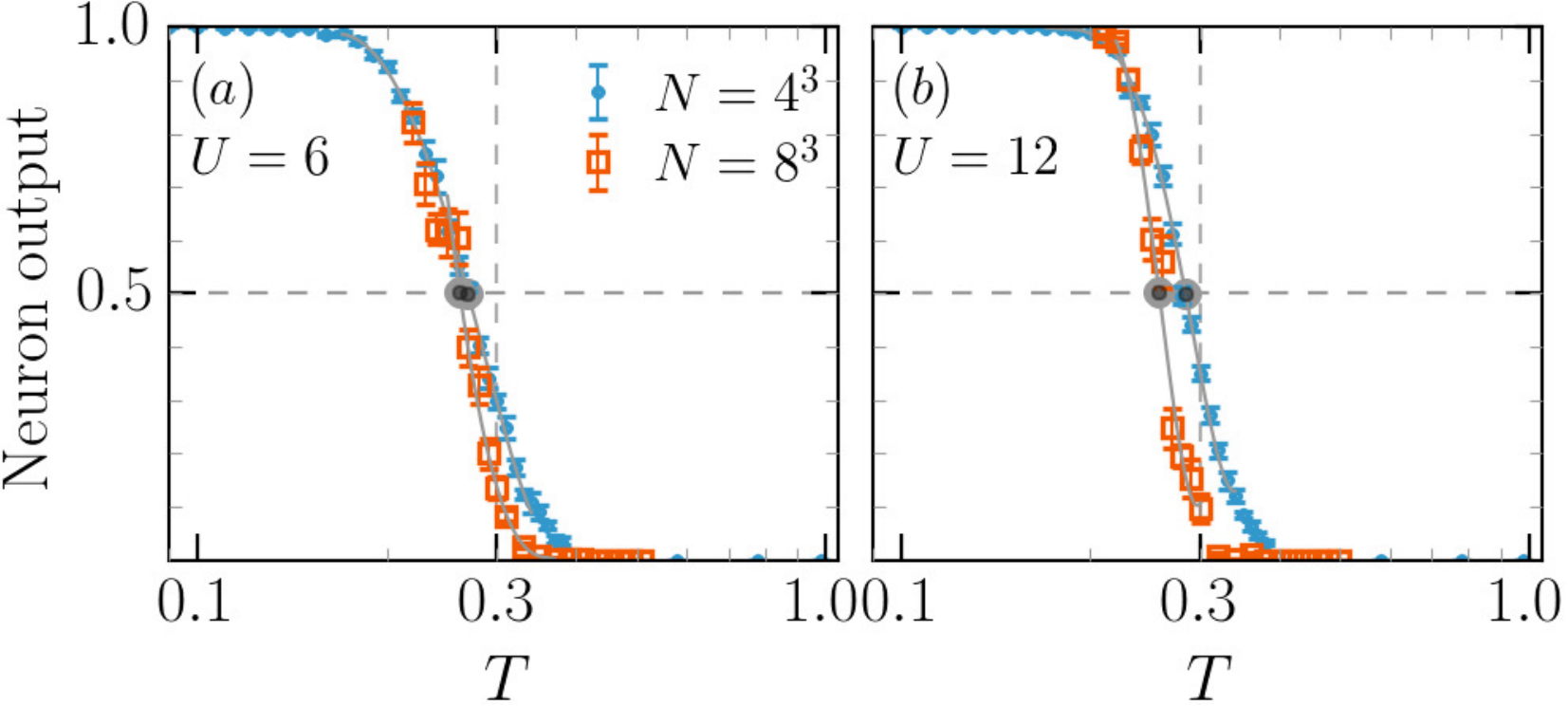}}
\caption{Average output of the neuron that is trained to be activated (return 1) in the ordered phase as a function of temperature at half filling. The network is maximally confused (average is 0.5) at the transition temperatures. We use 90 000 configurations for $N=4^3$ and 10 000 configurations for $N=8^3$. The grey solid lines are fits to data in the ranges shown. The vertical dashed lines indicate the estimates for $T_N$ in the thermodynamic limit (about 0.30 for both $U$ values) from other studies (see caption of Fig.~\ref{fig:TN}).
\vspace{-0.1in}
\label{fig:cross}}
\end{figure}

The predicted critical temperatures for other values of $U$ during the classification are taken as temperatures at which the network is maximally ``confused", i.e., when the average output crosses 0.5. Figure~\ref{fig:cross} shows the average output of the neuron, in the classification, that is trained to be activated, i.e., to return 1, in the ordered. Results are shown for the two system sizes and for $U=6$ and 12. We perform the classifications on 90 000 (10 000) configurations per temperature for $N=4^3$ ($N=8^3$). The data become noisier with a smaller number of configurations for the larger system and by increasing $U$, and so we use fits to a third-degree polynomial using data near the 0.5 crossing to better estimate $T_N$.

\begin{figure*}[t]
\centerline {\includegraphics*[width=7in]{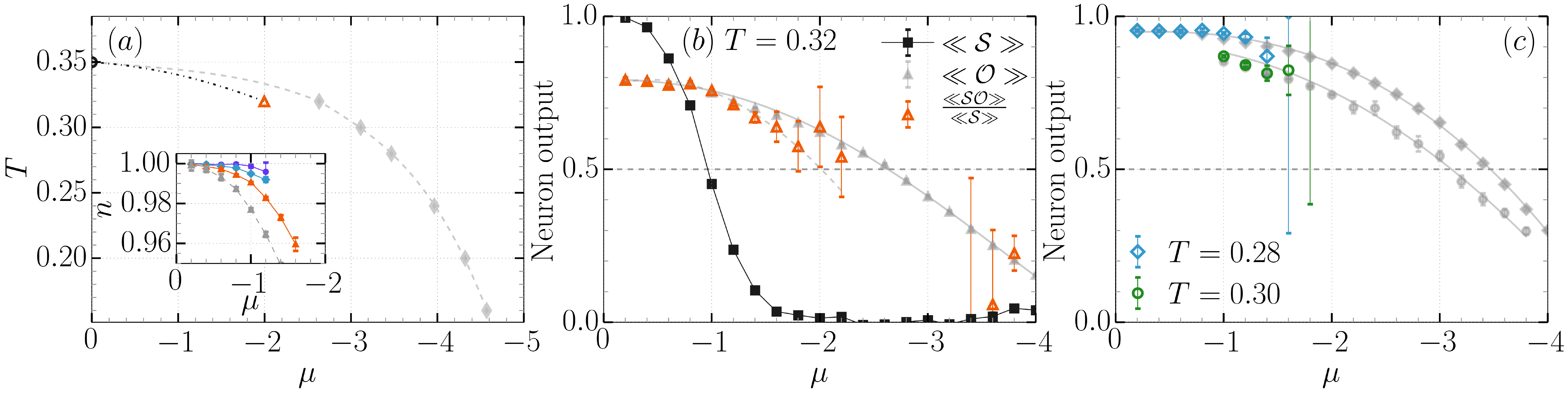}}
\caption{Phase transitions away from half filling. (a) Magnetic phase diagram of the 3D Hubbard model away from half filling for $U=9$. Empty red (filled grey) symbol(s) are estimates for the N\'{e}el temperatures, taking (not taking) the minus sign problem into account, obtained from a neural network that is trained to identify the phase at half filling. Lines are guides to the eye. Inset: Equations of state (density vs $\mu$) for $U=9$ at $T=0.24$, 0.28, and 0.32 (solid lines from top to bottom, respectively). The grey dashed line is calculated without taking the sign problem into account at $T=0.32$. (b) Average neuron output $\mathcal{O}$ calculated taking (without taking) the sign into account at $T=0.32$ (${\llangle \mathcal{S O} \rrangle}/{\llangle\mathcal{S} \rrangle}$ and $\llangle \mathcal{O} \rrangle$, respectively). See Appendix~$\hyperref[app:Appendix_A]{\text{A}}$ for details. The grey dotted line is a fit to the data near the 0.5 crossing point. (c) Same as in panel (b) but for $T=0.30$ and $0.28$.
\vspace{-0.1in}
\label{fig:away}}
\end{figure*}

The simultaneous use of configurations for $U=5$ and 16 is crucial in obtaining the results in Fig.~\ref{fig:TN}. We find that a network that is trained to distinguish phases only in the weak-coupling regime, e.g., for $U=5$, correctly predicts the trend in $T_N$ vs $U$ for systems in the weak-coupling regime but grossly overestimates it for systems in the intermediate- and strong-coupling regimes. Similarly, a network that is trained with configurations only for $U=16$ underestimates $T_N$ for any $U < 16$ (see the open symbols in Fig.~\ref{fig:TN}). Physically, this can be understood as follows: The transition to the AFM ordered phase is less sharp in the weak-coupling regime, with substantial double occupancy still present in the system above the transition. As a result, the network that is trained only in this regime tracks the onset of Mott physics, which moves to higher temperatures as the interaction strength increases, not the N\'{e}el transition. On the other hand, the network that is trained only in the strong-coupling regime tracks the onset of the region with significantly reduced double occupancy and stronger AFM correlations, which takes place at lower temperatures for smaller values of $U$. The competition between these two scenarios when training with both $U=5$ and $U=16$ data is encoded in our CNN and is key in obtaining reasonable $T_N$ for other $U$. 

Does the magnetically ordered phase survive if we dope the system away from half filling, and if so, what is the dependence of $T_N$ on doping? We try to answer these questions for $U=9$, where the transition temperature is highest at half filling. We train the network using spin configurations obtained for $U=9$ at half filling and then classify other configurations generated at fixed temperatures less than the half-filled value of $T_N$, but across the chemical potential axis. As soon as $\mu$ deviates from zero (the value corresponding to half filling), the weight of configurations in the DQMC can become negative, leading to the so-called ``sign problem"~\cite{e_loh_90,v_iglovikov_15}. We treat the neuron output as a (binary) physical observable, which, like spin correlations, can be written as a nonlinear function of the auxiliary spins (see Appendix~$\hyperref[app:Appendix_A]{\text{A}}$). Therefore, the expectation value is calculated in the conventional way by including the sign in the averaging and dividing by the average sign, typical for DQMC (see Appendix~$\hyperref[app:Appendix_A]{\text{A}}$). This procedure is valid as long as the average sign does not vanish. For completeness, in the following, we also show results in cases where we have ignored the sign problem and performed neuron output averages simply by using the absolute value of weights of the configurations.

The results are summarized in Fig.~\ref{fig:away}(a), where we show the location of the critical $\mu$, the onset of dominant AFM correlations, at $T=0.32$ (red triangle). Similarly to $T_N$ at half filling, the critical $\mu$ is estimated as the chemical potential where the expectation value of the neuron output crosses 0.5. To note the effect of the negative sign problem on our results, the grey filled points show the results if the sign is ignored during the calculation of the expectation value. The transition temperature remains nonzero at $n\ne 1$ but is expected to rapidly decrease by decreasing the density. The flat region at $\mu>-1.0$ is a direct consequence of the Mott physics setting in near half filling. At the temperatures we have access to, the Mott gap is not fully developed yet, and the density starts deviating from unity around $\mu= -1.0$. This is more clearly seen in the equations of state shown in the inset of Fig.~\ref{fig:away}(a) at $T=0.24$, 0.28, and 0.32.  

In Figs.~\ref{fig:away}(b) and~\ref{fig:away}(c), we show the average neuron output taking and without taking the sign problem into account vs $\mu$ at $T=0.32$, $0.30$, and $0.28$. In Fig.~\ref{fig:away}(b), we also show the average sign in the DQMC at $T=0.32$. As the latter approaches zero around $\mu=-2$, the accuracy in the expectation value of the neuron output is largely compromised. Although it appears that the sign problem does not have a considerable effect on the value of our observable for this particular magnetic phase transition, we find significant differences between the results obtained taking and without taking the sign into account at $T=0.32$ starting at $\mu=-1.4$. The latter can call into question a recent suggestion that the sign problem can be circumvented using neural networks by essentially ignoring the sign in training or classifications~\cite{p_broecker_16}. We fit the data with error bars to a third-degree polynomial for $\mu\ge -2.2$ and deduce a critical $\mu$ of $-2.0$ for the 0.5 crossing, which is larger than $-2.6$ obtained by ignoring the sign problem. Note that here we are ``transfer learning" by employing a network that has been trained in the sign-problem-free parameter region and using it in the sign-problematic region, an approach that can be used in other machine learning applications where training in the sign-problematic region cannot be justified. At $T < 0.32$, the sign problem becomes more severe and prevents us from obtaining estimates for $T_N$ [see Fig.~\ref{fig:away}(c)].

The inset of Fig.~\ref{fig:away}(a) shows that ignoring the sign problem also leads to a smaller average density at a given $\mu$ (see the grey dashed line in the inset corresponding to $T=0.32$). This leads to a magnetic phase diagram in the more physical temperature-density space that very much depends on whether or not the sign has been properly taken into account. Our results suggest that, at $T=0.32$ and the critical $\mu=-2$, the density is less than $0.95$, consistent with recent results from a more conventional method~\cite{e_khatami_16}.

\section{conclusion}

We have utilized neural-network machine learning techniques to predict the onset of the finite-temperature magnetically ordered phases of the 3D Fermi-Hubbard model. We train a 3D CNN using auxiliary spin configurations for a range of temperatures around the transition temperature, sampled over during DQMC simulations of two systems in the weak- and strong-coupling regimes. We show that the trend in the N\'{e}el temperature of the half-filled model as a function of the interaction strength can be captured by using the trained network to classify configurations generated for other interaction strengths. We then train a network at half filling for $U=9$ and use it to predict the fate of the ordered phase as the system is doped away from half filling. We find that the instability persists in the latter region in close proximity to the commensurate filling and that not including the sign in the expectation value of the neuron outputs leads to different results.

\section*{Acknowledgements}

We acknowledge useful conversations with Rajiv R. P. Singh, Andy Millis, Simon Trebst, and Peter Broecker. K.C. and E.K. acknowledge support from the U.S. National Science Foundation under Grant No. DMR-1609560. R. M. acknowledges support from NSERC and the Canada Research Chair program. Additional support was provided by the Perimeter Institute for Theoretical Physics. Research at Perimeter Institute is supported by the Government of Canada through the Department of Innovation, Science and Economic Development Canada and by the Province of Ontario through the Ministry of Research, Innovation and Science.

\setcounter{equation}{0}
\setcounter{section}{1}
\renewcommand{\thesection}{\Alph{section}}
\renewcommand{\theequation}{\thesection \arabic{equation}}
\section*{APPENDIX \Alph{section}: DETERMINANT QUANTUM MONTE CARLO} \label{app:Appendix_A}

In the DQMC, the partition function $Z=\textrm{Tr}e^{-\beta H}$is expressed as a path integral by discretizing the inverse temperature $\beta$ into $\mathcal{L}$ slices of length $\Delta\tau$. The one-body (kinetic) and two-body (interaction) terms of the Hubbard Hamiltonian are then separated in each time slice, leading to a product of two exponentials: $Z=\textrm{Tr}(e^{-\Delta\tau K}e^{-\Delta\tau V})^\mathcal{L}+\mathcal{O}(\Delta\tau^2)$, where $K$ and $V$ are the kinetic and interaction parts of the Hamiltonian, respectively. Since the two terms do not commute, this process introduces a small controlled error of the order of $\Delta\tau^2$. The interaction terms at different times can then be written in terms of one-body (quadratic) fermion operators using the Hubbard-Stratonovich (HS) transformation

\begin{equation}
e^{-U\Delta\tau	n_{\uparrow}n_{\downarrow}}= \frac{1}{2}e^{-U\Delta\tau (n_{\uparrow}+n_{\downarrow})/2}
\sum_{s=\pm 1}e^{-s\lambda (n_{\uparrow}-n_{\downarrow})},
\end{equation}

where $\cosh \lambda= e^{U\Delta\tau /2}$, at the expense of introducing a sum over the field of auxiliary (spin) variables like $s$ in a $D + 1$-dimensional space ($D$ spatial and 1 imaginary time). The fermionic degrees of freedom in the quadratic form can be integrated out analytically, resulting in the product of two determinants, one for each fermion species, and leaving behind the sum over the $2^{N\mathcal{L}}$ configurations of the auxiliary spins, where $N$ is the spatial size of the system. Hence, the partition function can be expressed as

\begin{equation}
Z\propto \sum_{\{s_{i,l}\}} \det M_{\uparrow}(\{s_{i,l}\})\det M_{\downarrow}(\{s_{i,l}\}),
\end{equation}

where $M_{\sigma}$ is an $N\times N$ matrix that depends on the spin configuration $\{s_{i,l}\}$, and $i$ and $l$ represent the spatial and time indices.

Observables are estimated by important sampling over the configurations, which is performed, e.g., using the Metropolis algorithm, with the product of determinants as the probability, to accept or reject proposed local changes to the field. However, the determinants are costly to update and, other than in a few special cases, can take different signs at low temperatures, leading to the infamous ``sign problem"~\cite{e_loh_90,v_iglovikov_15}. At half filling, the two determinants have the same sign by symmetry, and thus, there is no sign problem. Away from half filling, one can take the absolute value of the product of determinants as the new probability in the Metropolis algorithm, but we have to treat the sign with the observable explicitly. The expectation value of a physical observable, $O$, can be calculated using

\begin{equation}
\left<O \right>=\frac{\llangle \mathcal{S} O \rrangle}{\llangle\mathcal{S} \rrangle},
\end{equation}

where $\mathcal{S}$ denotes the sign of the product of determinants and $\llangle \cdot \cdot \rrangle$ represents the Monte Carlo average using the absolute value of the product of determinants as the probability. Here, we take the neuron outputs as observables. We use jackknife resampling to estimate the error bars.

The particular decoupling scheme we have chosen is especially useful for inferring magnetic ordering by the machine purely based on the auxiliary fields. This is because the correlations between fermionic spins $S$ can be written in terms of auxiliary spins~\cite{j_hirsch_83,j_hirsch_86b}:

\begin{equation}
\left < S_i(\tau)S_j(0)\right>= (1-e^{-\Delta\tau U})^{-1}\left<s_i(\tau) s_j(0) \right>,
\end{equation}

(not valid when $\tau=0$ and $i=j$). Therefore, the auxiliary field variables are effectively representing the fermion spins with the same correlations in space and time. The decoupling can be done using other variables that couple to density or pairing operators, which can be more helpful in detecting other phases, including charge density wave or superconductivity. We do not explore those possibilities here.

We use the quantum electron simulation toolbox (QUEST)~\cite{quest} for our DQMC simulations. The toolbox, with minimal modifications, allows for outputting of the HS field configurations at arbitrary intervals during the simulations. We use 1000 (500) warmup sweeps and between 10 000 and 100 000 (1000 and 10 000) measurement sweeps for $N=4^3$ ($N=8^3$) in each run. We repeat runs with different random number seeds to obtain the desired number of configurations. In most cases, the HS field configuration is written into a file after every 10 sweeps, the same frequency typically used to perform physical measurements. For training, we choose the temperature grid for every value of $U$ so that we have the same number of temperature points above and below the expected $T_N$. The temperatures are chosen in a nonuniform grid that can extend from 0.03 to about 3.0 depending on $U$.

The discretization of the imaginary time interval $[0 - \beta]$ ($\beta=1/T$) in the DQMC introduces a systematic Trotter error, which scales like $\Delta\tau^2$, where $\Delta\tau=\beta/\mathcal{L}$ and $\mathcal{L}$ is the number of time slices. Thus, in principle, one has to choose a variable $\mathcal{L}$ at different temperatures so that the error in physical observables is of the same order. However, such configurations, with variable size of the fourth dimension, are not useful in the training of a network whose size of the input layer has to be kept fixed. Hence, we choose a large fixed $\mathcal{L}=200$ for all $\beta$ such that $\Delta\tau$ remains below the threshold value of $(8U)^{-1/2}$, where the Trotter error can be neglected, for the largest $U$ values and the lowest temperatures considered here. Moreover, we are not interested in calculating physical observables; rather, we are interested in locating the critical points at which the correlations of the system, both in the spatial and in the imaginary time dimensions, diverge; thus, the discretization errors can essentially be ignored.

\setcounter{equation}{0}
\setcounter{section}{2}
\section*{APPENDIX \Alph{section}: CONVOLUTIONAL NEURAL NETWORK} \label{app:Appendix_B}

\renewcommand{\thesubsection}{\arabic{subsection}}
\subsection{Case of $N=4^3$}

Each auxiliary field configuration is fed into a 3D CNN, with each imaginary time slice encoded as a different filter channel (input block with $4^3$ neurons---see the input layer in Fig.~\ref{fig:network}). Each block from the $\mathcal{L}$ imaginary time slices is immediately connected to 32 hidden volumetric feature maps (blocks in the first hidden layer) for feature extraction, for the network to detect a collection of unique features.

\begin{figure}
\centerline {\includegraphics*[width=2.5in]{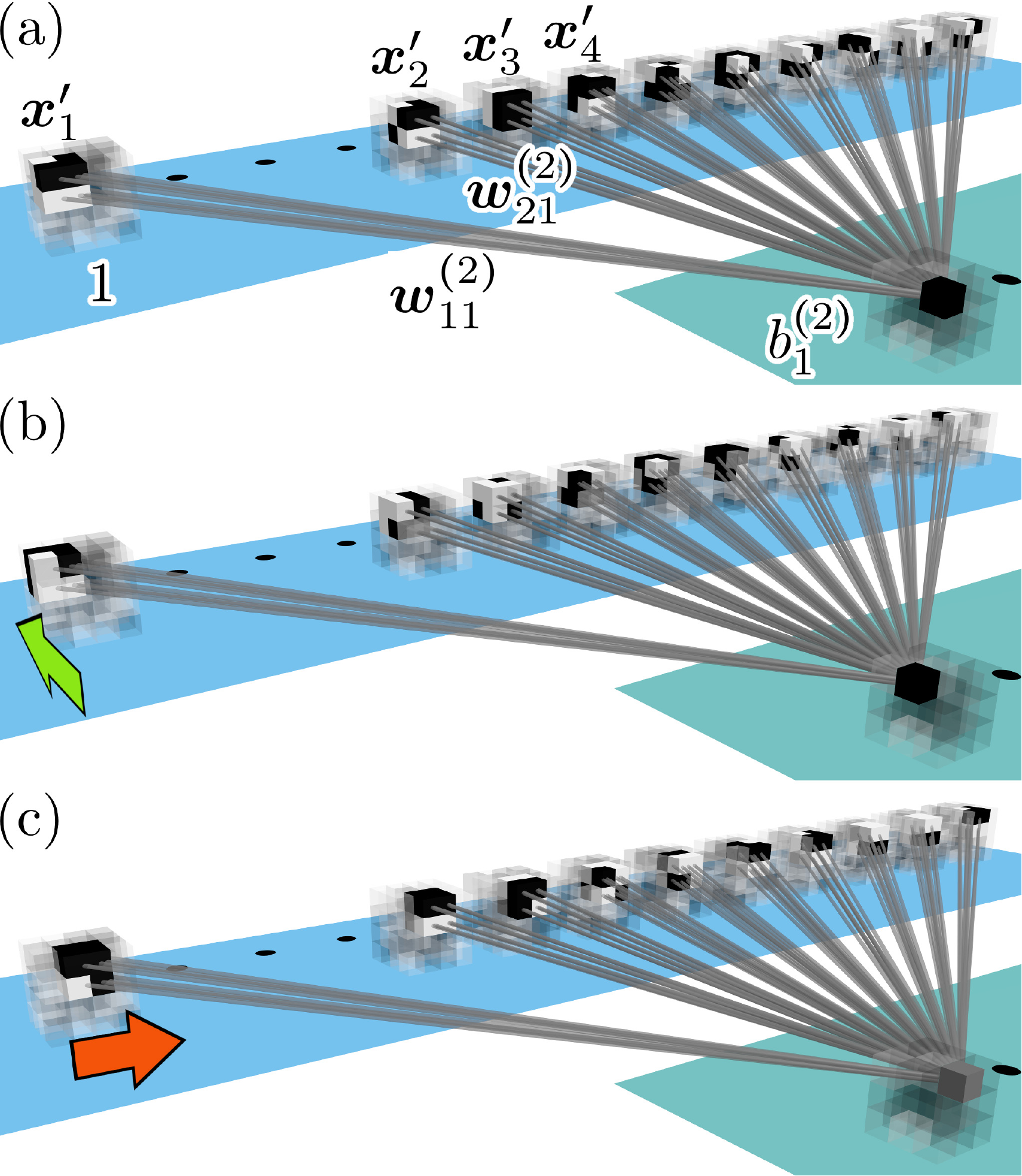}}
\caption{Construction of a volumetric feature map. To produce the highlighted neuron in the volumetric feature map, from the input, a shared filter $\boldsymbol{w}_{l1}^{(2)}$ is convolved with a $2\times2\times2$ local receptive cube $\boldsymbol{x}_{l}'$ across all the inputs in the imaginary time slices, adding a bias offset $b_{1}^{(2)}$, before being activated using ReLU. Sliding the $2\times2\times2$ cube in each of the spatial dimension while repeating the convolutional procedure completes a volumetric feature map. The spatial size of the volumetric feature map can be preserved by using zero-padding.
\label{fig:222}}
\end{figure}

To produce the volumetric feature maps in the first hidden layer, we convolve a shared filter (also known as a kernel) $\boldsymbol{w}_{lm}^{(2)}$ with a $2 \times 2 \times 2$ local receptive cube that sweeps each input block, and we offset the result with a shared bias $b_{m}^{(2)}$. The kernel $\boldsymbol{w}_{lm}^{(2)}$ can be thought of as a $2^3$ tensor, where the superscript denotes the layer it corresponds to, $l \in [1,\mathcal{L}]$ is the input block label, which is the same as the imaginary time index, and $m\in [1,32]$ labels the blocks in the first hidden layer. The convolution can be written as a tensor operation $z_{lmn}=\boldsymbol{w}^{(2)}_{lm} \boldsymbol{x}_{ln} + b^{(2)}_m$, where $\boldsymbol{x}_{ln}$ represents the $n$th input as picked up by the local receptive cube as it sweeps the $l$th input block. The process is depicted in Fig.~\ref{fig:222}. Before $z$ is written in the corresponding element in the hidden layer, we pass it through a rectified linear unit (ReLU) $f(z)=\text{max}(0, z)$, serving as the neural activation function. 

Sliding the local receptive cube with a stride of one in each spatial dimension on the $\mathcal{L}$ input cubes completes the construction of a volumetric feature map. To ensure that the information around the borders does not deplete too quickly, especially with smaller input size, as more convolutions are performed in the next layers, we use zero padding to preserve the input size after each convolution; we pad each of the input cubes with zeros, increasing their size to $5 \times 5 \times 5$ so that the output volume has the same size as the original data, i.e., $4 \times 4 \times 4$.

As deeper networks are typically more powerful, we use two more convolutional layers with ReLUs. We employ 16 and 8 volumetric feature maps for the second and third hidden convolutional layers, respectively. We repeat the procedure described above to construct the feature maps in these layers.

The final feature extraction layer is connected to the first classification layer, where some judgement on the input is made (see Fig.~\ref{fig:network}). There are eight neurons in this layer. Each of the $8\times 4^3$ individual neurons from the third feature extraction layer is connected to each of the fully connected neurons in the first classification layer. Finally, each of the eight hidden neurons is connected to two output neurons, where the final judgement is made, using softmax as their activation function. One of the output neurons represents the ordered state $(T < T_{N})$, and the other represents the unordered state $(T > T_{N})$. The output of the softmax functions represents the likelihood of the input configuration belonging to each of the two categories. As such, the sum of outputs from the two neurons is always 1. The activated output neuron is taken to be the one with the highest probability.

During training, we use dropout regularization with a dropout rate of 0.5 on the eight fully connected neurons in order to mitigate overfitting. Namely, half of the eight fully connected neurons are chosen at random and temporarily deactivated. This forces the neurons to adapt to more robust features.

An exponentially decaying learning rate is also used to ensure that the network starts out learning rapidly and slows down the learning after the model is close to convergence. The learning rate is given by

\begin{equation}
    \eta=\eta_{0} \lambda^{\mathrm{training\ epoch}},
    \label{eq:learning_rate}
\end{equation}

\noindent
where the initial learning rate $\eta_{0}=10^{-3}$, and the decay rate $\lambda=0.925$. A complete training epoch is defined as when the network has stepped through the whole set of training data once.

We use a cross-entropy cost function given by~\cite{Nielsen2016}

\begin{equation}
    C=- \frac{1}{n_{\mathrm{td}}} \sum_{x} \sum_{i=1}^{2} [ y_{i} \ln a_{i} + (1-y_{i}) \ln(1-a_{i}) ],
    \label{eq:cross-entropy}
\end{equation}

\noindent
where $x$ is the input data, $n_{\mathrm{td}}$ is the number of training data, $y_{i}$ is the desired output, and $a_{i}$ is the network output.

\subsection{Case of $N=8^3$}

The same procedure described above for the $N=4^3$ system is used for $N=8^3$. However, to keep the computational cost of the optimizations manageable, for $N=8^3$, we perform convolutions without zero padding. We also use four hidden-feature extraction layers in this case. To prevent overshooting the optimal model, we use a more aggressive decay rate $\lambda=0.875$.

We find that handpicking the number of feature maps and fully connected neurons, similar to what we do for the case of $N=4^3$, yields no learning. Thus, to avoid a costly hyperparameter grid search, where hyperparameter refers to the number of feature maps or number of neurons in the classification layer, we perform a Monte Carlo sampling for 20 search iterations in the hyperparameter space. We constrain the number of feature maps and fully connected neurons to the interval $[8,128]$. The optimized hyperparameters we found (the set that yields the largest training accuracy) are $n^{(2)}=54$, $n^{(3)}=26$, $n^{(4)}=14$, $n^{(5)}=18$, and $n^{(6)}=64$, where $n^{(2)}$ through $n^{(5)}$ are the number of feature maps in the feature extraction layers and $n^{(6)}$ is the number of neurons in the first classification layer.

\setcounter{section}{3}
\section*{APPENDIX \Alph{section}: TRAINING THE CNN} \label{app:Appendix_C}

\begin{figure}[b]
\centerline {\includegraphics*[width=3.3in]{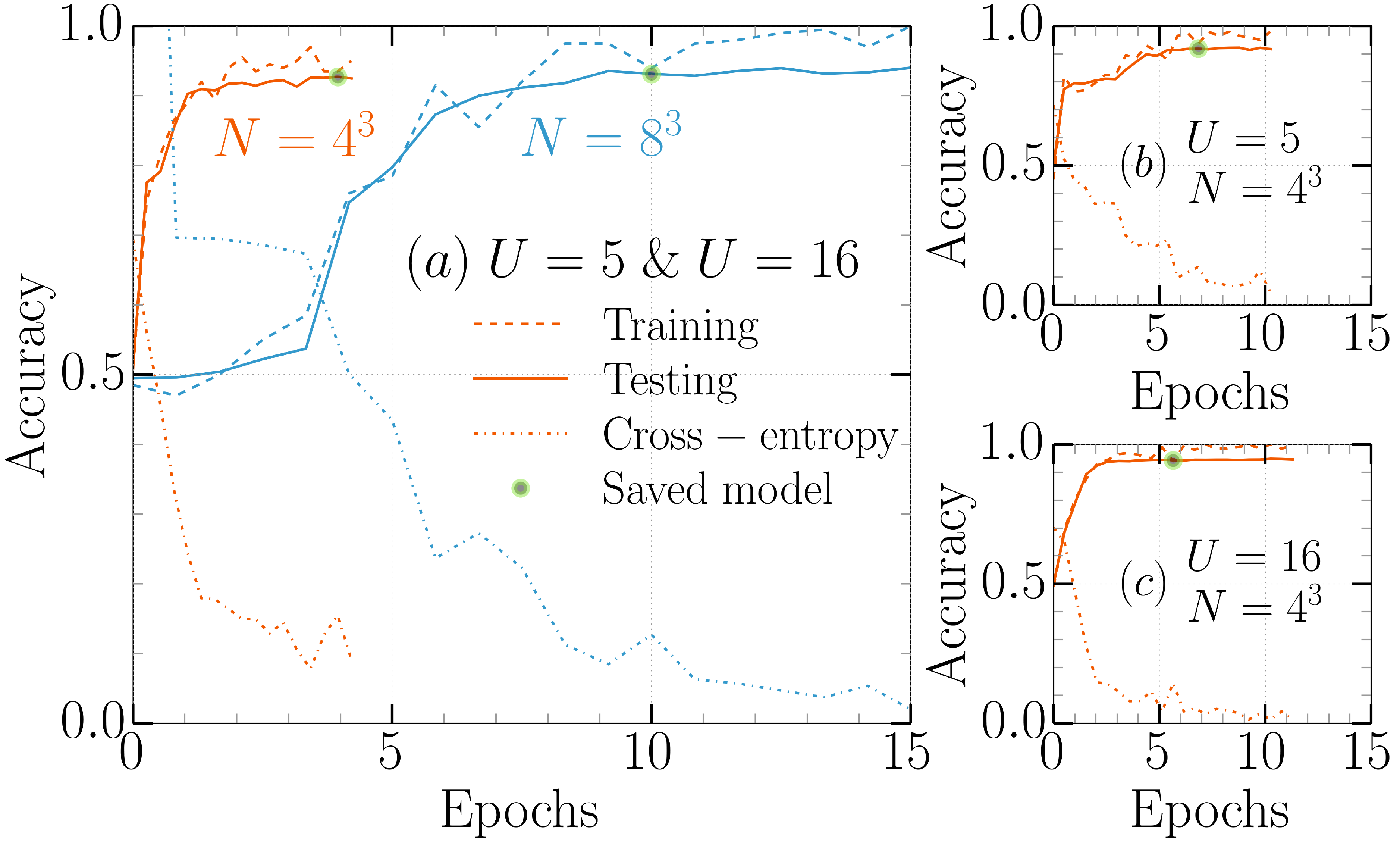}}
\caption{Progress of the training process. Training accuracy for (a) simultaneous training with $U=5$ and 16, (b) $U=5$ alone, and (c) $U=16$ alone vs epochs. The location of last saved model is shown as a green circle in each panel. The testing accuracies for saved models are 92.7\% (93.2\%) for $N=4^3$ ($N=8^3$) in (a), 91.9\% in (b), and 94.3\% in (c).
\label{fig:train}}
\end{figure}

Figure~\ref{fig:train} shows the training progress of our networks and the evolution of the cost function in different cases vs training epochs. In a training epoch, we go through the entire set of shuffled training data once through iterations that involve batch sizes of 200. Here, 85\% of the input data was used for training and 15\% for testing. Test data are used as a gauge for overfitting (overtraining). The network overfits to the training data when the training accuracy continues to improve, but the testing accuracy stalls. We stop the training once we detect overfitting. The criterion is that the training accuracy and testing accuracy should not deviate from each other for more than 10 successive training evaluations.

\begin{figure*}[t]
\centerline {\includegraphics*[width=7in]{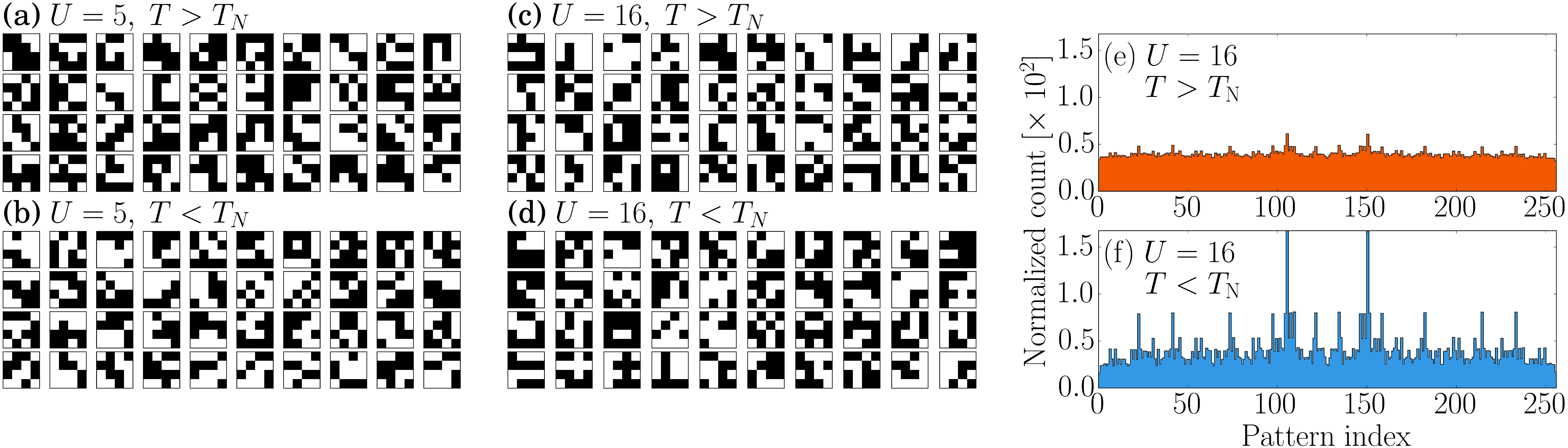}}
\caption{(a)--(d) Typical auxiliary field patterns above and below $T_N$ used as input to the CNN for the $N=4^3$ system. In each panel, the four rows correspond to the four layers in the $z$ direction. We show the field only in the first 10 imaginary time slices (left to right) for each case. (e,f) Normalized count of the 256 patterns within $2\times 2\times 2$ cubes as detected in the raw configurations for $U=16$ below and above $T_N$. The two largest peaks in (f) correspond to the two AFM patterns.
\label{fig:fields}}
\end{figure*}

We save the model (weights and biases) every time (i) the difference between the training accuracy and testing accuracy is less than 2.5\%, (ii) the training accuracy is greater than the testing accuracy, and (iii) the current testing accuracy is better than the last recorded testing accuracy. The locations of the last saved models are shown as green circles for each case in Fig.~\ref{fig:train}.

\setcounter{equation}{0}
\setcounter{section}{4}
\section*{APPENDIX \Alph{section}: WHAT DOES THE MACHINE LEARN?} \label{app:Appendix_D}

One may wonder what features are learned by the CNN, as encoded in the weights and biases. Before addressing this question, we make one observation. Unlike in cases of image or sound recognition, the information put into our machine (auxiliary spin configurations) consist of binary numbers; namely, each input neuron takes 1 or -1 as the value. Therefore, the ``features" extracted by the eight-site receptive cube will be no more than $2^8=256$ distinct patterns of 1's and -1's on a $2\times 2 \times 2$ cube. In Figs.~\ref{fig:fields}(a)-\ref{fig:fields}(d), we show a typical auxiliary field for the first 10 time slices from DQMC simulations of the system with $U=5$ and $U=16$ both below and above $T_{N}$. One can see that, due to quantum fluctuations, a clear AFM pattern that can be discerned does not emerge in the 2D images. However, plotting the histogram of all 256 patterns in a sample set for $U=16$ in Figs.~\ref{fig:fields}(a)-\ref{fig:fields}(d) reveals that the AFM pattern is in fact the dominant one at $T < T_N$.

\begin{figure}[b]
\centerline {\includegraphics*[width=3.in]{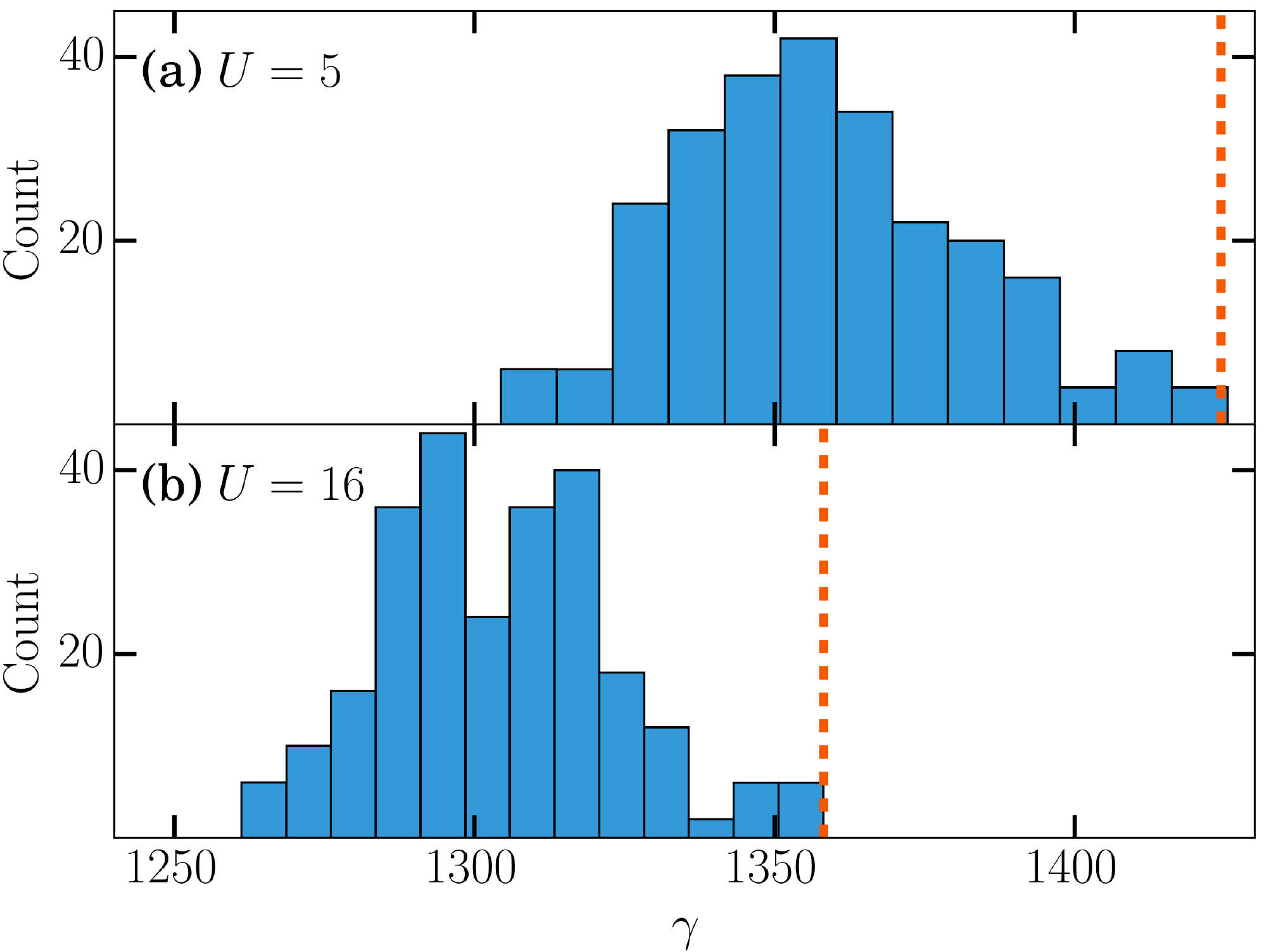}}
\caption{Histogram of the overlap between the learned features and all the possible ordering patterns picked up by the $2\times 2\times 2$ receptive cube. Top and bottom panels correspond to cases with $U=5$ and $U=16$, respectively. The vertical dashed (red) line marks the value for the perfect AFM ordering pattern.
\label{fig:hist}}
\end{figure}

We focus on the first feature extraction layer immediately after the input layer. What has been encoded in the weights and biases in that layer should be a good indication of what the CNN is looking for in all the feature extraction layers. Specifically, we would like to know to what extent the saved $\boldsymbol{w}^{(2)}_{lm}$ correlate with each of the 256 ordering patterns. To find out, we convolve each of the ordering patterns with each $\boldsymbol{w}^{(2)}_{lm}$. We introduce the following overlap function:

\begin{equation}
\gamma_{\boldsymbol{x}}=\sum_{lm} |\boldsymbol{w}^{(2)}_{lm} \boldsymbol{x} |
\end{equation}
where each $\boldsymbol{x}$ is a $2\times 2\times 2$ tensor with 1's and -1's as elements corresponding to one of the 256 ordering patterns on the eight-site cube. We have safely ignored the biases as they introduce a uniform shift and do not pertain to local correlations. We take the absolute value of the tensor product before performing the sums since we expect the spin inversion symmetry to have been encoded in the CNN to a good extent during the training; i.e., $\sum_{lm} \boldsymbol{w}^{(2)}_{lm} \boldsymbol{x}$ orders of magnitude smaller than $\gamma_{\bf x}$ for any of the ordering patterns, something we confirm with our network. The dominant features, as seen by the CNN, are those with the largest $\gamma_{\bf x}$. Here, $\gamma_{\bf x}$ plays a role analogous to the order parameter for each of the ordering patterns. For example, ${\bf w} {\bf x}^{\textrm{AFM}}$, where ${\bf x}^{\textrm{AFM}}$ corresponds to the perfect AFM pattern, would be the staggered ``magnetization" of ${\bf w}$.

In Fig.~\ref{fig:hist}, we show histograms of $\gamma_{\bf x}$ for the 256 possibilities for ${\bf x}$ at $U=5$ and $U=16$, respectively. The ones in the large-$\gamma_{\bf x}$ tail of the distribution are the dominant patterns. The red vertical line denotes the magnitude of the convolution (overlap) with the two degenerate AFM patterns. We find that for both a CNN trained by the $U=5$ data and one trained by the $U=16$ data, the AFM pattern is clearly the dominant pattern learned by the network.



%

\end{document}